\documentclass[aps,prd,twocolumn,superscriptaddress,showpacs,showkeys]{revtex4-1}
\usepackage{amsmath}
\usepackage{amsfonts}

\begin{document}
\title{Constraining the Existence of Magnetic Monopoles by Dirac-Dual Electric Charge Renormalization Effect Under the Planck Scale Limit}
\author{Yanbin Deng$^{1,3}$, Changyu Huang$^{2}$, Yong-Chang Huang$^{1,4}$}
\affiliation{$^{}$Institute of Theoretical Physics, Beijing University of Technology, Beijing 100124, China \\$^{2}$Department of Physics, Purdue University, W. Lafayette, IN 47907-2036, USA \\$^{3}$Department of Physics, Baylor University, Waco,  Texas 76706, USA\\ $^{4}$CCAST(WorldLab.), P.O. Box 8730, Beijing 100080, China}
\date{\today}

\begin{abstract}
It was suggested by dimensional analysis that there exists a limit called the Planck energy scale coming close to which the gravitational effects of physical processes would inflate and struggle for equal rights so as to spoil the validity of pure non-gravitational physical theories that governed well below the Planck energy. Near the Planck scale, the Planck charges, Planck currents, or Planck parameters can be defined and assigned to physical quantities such as the single particle electric charge and magnetic charge as the ceiling value obeyed by the low energy ordinary physics. The Dirac electric-magnetic charge quantization relation as one form of electric-magnetic duality dictates that, the present low value electric charge corresponds to a huge magnetic charge value already passed the Planck limit so as to render theories of magnetic monopoles into the strong coupling regime, and vice versa, that small and tractable magnetic charge values correspond to huge electric charge values. It suggests that for theoretic models in which the renormalization group equation provides rapid growth for the running electric coupling constant, it is easier for the dual magnetic monopoles to emerge at lower energy scales. Allowing charges to vary with the Dirac electric-magnetic charge quantization relation while keeping values under the Planck limit informs that the magnetic charge value drops below the Planck ceiling value into the manageable region when the electric coupling constant grows to one fourth at a model dependent energy scale, and continues dropping towards half the value of the Planck magnetic charge as the electric coupling constant continues growing at the model dependent rate towards one near Planck energy scale.
\end{abstract}
%\pacs{12.15.Ji, 13.26.He}
\maketitle

\section{Introduction}
%\begin{center}
%\textbf{I. Introduction}
%\end{center}

Magnetic monopoles have been wished for by physicists due to a preference for detailed duality between electricity and magnetism.\cite{Wenwitten, Sen, SEIBERG, Gauntlett} Enthusiastic huntings for magnetic monopoles have not been defeated even though they have been evading experimental investigations for centuries.\cite{Milton, Super每Kamiokande, IceCubeCollaboration} Among the early important findings regarding magnetic monopoles in modern physics was Dirac's discovery of the electric-magnetic charge quantization relation.\cite{Dirac} It was well before people understood that both the electric and magnetic charges can vary with energy scale as a result of running coupling constants governed by the renormaliation gruop equation of their quantum field theories.\cite{Schwinger, Blagojevic, Zwanziger, LaperashviliNielsen} The Dirac magnetic monopole was theoretically augmented as a topological soliton solution to the 't Hooft-Polyakov monopole, which contains Dirac monopole as the far field limit.\cite{tHooft, Polyakov} Various attempts of grand unified theories(GUTs) revived the possibility for the existence of magnetic monopoles under situations like in the early universe. Magnetic monopoles have been found common products in GUT symmetry breaking processes down to Standard Model particle physics.\cite{Coleman, GODDARDNUYTSOLIVE, MONTONENOLIVE} As long as the GUT symmetry breaking process took place sometime before the end of cosmic inflation, a short period of rapid expansion of the very early universe, the primodial magnetic monopoles would be stretched over a large scale of the universe by the end of inflation so as to leave us very sparse trace of their appearance and existence.\cite{Preskill, GuthTye, Guth, AlbrechtSteinhardt} As a result, in spite of the possible abundance of primordial magnetic monopoles, their degree of scarcity in the present universe became adjustable by the energy scale of the GUT symmetry breaking. The earlier the GUT symmetry breaking, the magnetic monopoles would be more washed out by cosmic inflation. The lower energy scale for the GUT symmetry breaking, the better chance there can be for magnetic monopoles to escape the inflationary dilution and better hope will remain for the present magnetic monopole existence. It is an interesting topic still under active studies.\cite{Romero, LeeNair, Murayama, Sakai, Garattini, GhoshSingh}

Typical grand unification theories set the scenario at the energy scale many orders higher than that with the current collider experiments, say of the order $\sim 10^{16}GeV$. It is not so far from such an energy scale, called the Planck scale near which it is believed that a unified treatment of a quantum theory of gravity with the quantum theory of matter fields must be required. This has been the purpose of a laborious theoretical construction for around half a century yet still not fulfilled. Several provisional strategies, such as superstring theory and loop quantum gravity theory have been formulated after prolonged attempts. The final quantum theory of gravity is expected to be drastically different from the ordinary field theories since the straightforwardly quantized theory of general relativity was shown to be plagued with the non-renormalizable problem. It helps to define a group of physical parameters as the anchoring points of Planck scale, to together delineate the border of quantum gravity. The energy scale as a major controller of the strength of physical effects, stands out as the major threshold indicator to the border of quantum gravity: the Planck mass, or the Planck scale energy. Certain specific values to a set of physical quantities serve together to mark the boundary for the sphere of validity beyond which ordinary physical laws surrender their power to a yet unknown quantum theory of gravity.

The article first surveys the definition of a set of Planck parameters. The Planck mass, or Planck energy is the controlling threshold when the strength of gravitational effects inflate and spoil the validity of low energy pure non-gravitational physical theories. Comparable strength of electric Coulomb interaction and magnetic Coulomb-like interaction correspond to the definition of the Planck electric charge and Planck magnetic monopole. The concept similarly can be extended to define the Planck electric current and Planck magnetic current. The ultra high strength of interactions denoted by the Planck mass, Planck electric charge and Planck magnetic charge can be deemed as the ceiling value for the low energy ordinary physics. This becomes a constraint on the existence of magnetic monopoles.

The coexistence of an electrically charged particle and a magnetically charged particle, or magnetic monopole is allowable if the corresponding electric charge and magnetic charge satisfy the so-called Dirac electric-magnetic charge quantization relation.\cite{Dirac} Given that both these charges can vary in value, the Dirac relation dictates that the present low value electric charge corresponds to a huge magnetic charge value--already passed the Planck limit--only accessible to theories of strong couplings, and vice versa, that small and tractable magnetic charge values correspond to huge electric charge values. The way for the electric and magnetic charge values to vary is through the renormalization group flow of the quantum theory of their respective fields.

Treating the Dirac charge quantization relation as one form of electric-magnetic duality, it brings about one significant convenience: the unfamiliar magnetic charge renormalization can be traded for the familiar electric charge renormalization. An immediate finding follows that: at low energy scales when the electric coupling constant is small, the dual magnetic monopole is of huge magnitude even beyond the Planck limit and is intangible; when the electric coupling constant reaches one fourth with increased energy scales, the dual magnetic monopole begins to drop across the Planck limit to become manageable; and for further increased electric coupling constant with even higher energy scales, the dual magnetic monopole further drops toward a lower bound of half the Planck magnetic charge limit.

The article points out at last that the specific behaviour of the renormalization group equation of the electric coupling constant is model dependent. The general tendency is similar, but the specific behaviour depends on whether quantum electrodynamics, or $SU(2)\times SU(1)$ electro-weak unification theory, or supersymmetric/non-supersymmetric grand unification theories is employed. It concludes that for theoretic models in which the renormalization group equation provides rapid growth for the running electric coupling constant, it is easier for the dual magnetic monopoles to emerge at lower energy scales. This provides a way to distinguish models as whether be in favour of or to disfavour the existence of magnetic monopoles.

\section{Planck Charges, Planck Currents as Limits of Ordinary Physics}
%\textbf{II. Planck Charges as Mass, Electric Charge and Magnetic Monopole}

For later convenience, we first do some reviews on Planck charges.\cite{Wesson, Pavsic} Einstein's general relativity theory of gravity is thought to be the approximation of some quantum gravity theory necessary at ultra high spacetime curvature and energy scale while Newtonian gravitation is the weak field approximation to Einstein's theory of general relativity. Traditionally, through dimensional analysis, the deterministic parameters can be defined through Newtonian gravitation relation for the characterization of the so called Planck scale.
\begin{equation}
F=G\frac{m_{1}m_{2}}{r^{2}}.
\end{equation}
Expressing the dimension of Newtonian constant $G$ in terms of the fundamental physical units,
\begin{equation}
\lbrack G]=[F]\frac{L^{2}}{M^{2}}=\frac{ML}{T^{2}}\frac{L^{2}}{M^{2}}=\frac{L^{3}}{MT^{2}}.
\end{equation}
Requiring the three fundamental constants $G$, $\hbar $ ($[\hbar ]=\frac{ML^{2}}{T}$), and $c$ ($[c]=\frac{L}{T}$) be unit valued when expressed in terms of another three fundamental constants, the Planck length $l_{P}$, the Planck time $t_{P}$ and the Planck mass $m_{P}$:
\begin{eqnarray}
G &&=1 \cdot \frac{l_{P}^{3}}{m_{P}t_{P}^{2}}\approx 6.67408\times 10^{-11}\frac{m^{3}}{kg\cdot s^{2}}, \nonumber\\
\hbar &&=1\cdot \frac{m_{P}l_{P}^{2}}{t_{P}}\approx 1.054572\times 10^{-34}\frac{kg\cdot
m^{2}}{s}, \nonumber \\
c &&= \frac{l_{P}}{t_{P}}\approx 2.99792458\times 10^{8}\frac{m}{s},
\end{eqnarray}
would solve us $l_{P}$, $t_{P}$ and $m_{P}$ in terms of $G$, $\hbar $, $c $ with their numerical values,
\begin{eqnarray}
m_{P}&&=\sqrt{\frac{\hbar c}{G}}\approx 2.17651\times 10^{-8}kg\approx 1.2209\times 10^{19}
\frac{GeV}{c^{2}}, \nonumber\\
l_{P}&&=\sqrt{\frac{G\hbar }{c^{3}}}\approx 1.61620\times 10^{-35}m, \nonumber\\
t_{P}&&=\frac{l_{P}}{c}=\sqrt{\frac{G\hbar }{c^{5}}}\approx 5.39106\times 10^{-44}s.
\end{eqnarray}

To make some qualitative sense of the concept, consider the equivalence between a particle and a black hole. The Planck mass is a ultra large value as the mass or energy for a single particle, but a ultra small value as the mass for a black hole. A black hole has as its characteristic length the Schwarzschild radius($r_{S}=\frac{2Gm}{c^{2}}$) near which the gravitational effects of the black hole becomes violent, while a particle relies on the reduced Compton wavelength($\lambda =\frac{\hbar }{mc}$) as a characteristic length scale of salient quantum effects. The Compton wavelength is inversely proportional to, but the Schwarzschild radius proportional to the mass/energy of the particle-black hole common body. For mass values much smaller than the Planck mass, the Compton length is much greater than the Schwarzschild radius, and the common body shows up as a quantum particle to external observers. For mass values much grater than the Planck mass, the Schwarzschild radius is much greater than the Compton length, and the common body shows up as a macroscopic black hole to external observers. It is a special case that near the Planck mass/energy, the Compton length and the Schwarzschild radius are close to each other at the value of the Planck length, so that the gravitational and the quantum effects of the particle-black hole common body are both significant, and the common body shows both as a quantum particle and a black hole--two facets indivisible. The quantum particle physics and gravitation are no longer disconnected with each other under this situation. Also note that the Planck time is the characterization of how rapidly the relevant physical processes are bursting at the Planck scale. The Planck scale physics in Nature is thought to occur at the beginning of the Big Bang and inside black holes.

Limits exist for various physical processes to stop them from running into the dangerous zone of quantum gravity physics. Bringing in the concept of Planck electric charge and Planck magnetic monopole marks fuller delineation of the Planck scale. Based on the electric Coulomb's law, in its scalar form,
\begin{equation}
F=\frac{1}{4\pi \epsilon _{0}}\frac{Q_{1}Q_{2}}{r^{2}}=k_{e}\frac{Q_{1}Q_{2}}{r^{2}},
\end{equation}
the requirement that the Coulomb's constant $k_{e}$ assumes unit value in Planck units renders the definition of the Planck electric charge.
\begin{equation}
\lbrack k_{e}]=[F]\cdot \frac{L^{2}}{Q^{2}}=\frac{ML}{T^{2}}\cdot \frac{L^{2}}{Q^{2}}=\frac{ML^{3}}{T^{2}Q^{2}},
\end{equation}
\begin{equation}
k_{e}=1\cdot \frac{m_{P}l_{P}^{3}}{t_{P}^{2}Q_{P}^{2}}\approx 8.987552\times 10^{9}N\cdot m^{2}\cdot C^{-2}.
\end{equation}
Employing the definition of the Planck length $l_{P}$, the Planck time $t_{P}$ and the Planck mass $m_{P}$ obtained above, the Planck electric charge is thus deduced:
\begin{equation}
Q_{P}=\sqrt{\frac{\hbar c}{k_{e}}}\approx 1.875546\times 10^{-18}C\approx 11.706\cdot e.
\end{equation}
Here $e$ is the unit electron charge measured at the present low energy experiment status.

Applying the above definition, the electric fine structure constant is related to the ratio of the low energy electron charge value to the Planck electric charge,
\begin{equation}
\alpha =\frac{e^{2}}{4\pi \epsilon _{0}\hbar c}=\frac{k_{e}}{\hbar c}\cdot e^{2}=\left( \frac{e}{Q_{P}}\right) ^{2}\approx \frac{1}{137.036}.
\end{equation}
The strength of the electric interaction, denoted by the electric fine structure constant, would scale up with the energy scale according to the renormalization group equation. Gliding over a large span of energy scales, the effects of renormalization group flow of quantum fields pushes up the electric fine structure constant towards the unit value when the elementary electron charge increases to the Planck electric charge,
\begin{equation}
\alpha |_{P}=\left( \frac{e}{Q_{P}}\right) ^{2}\Big|_{P}\rightarrow \left( \frac{Q_{P}}{Q_{P}}\right) ^{2}= 1.
\end{equation}

The validity of the ordinary perturbative quantum field theory would break down as the coupling constant approaches to unit value at some energy scale. The energy scale for this to happen is some ultra high, and can be assumed to be close to the Planck mass/energy before embarking on detailed numeric analysis, for regular behaviour field theories. The opposite is true for theories with asymptotic freedom.

If a strict duality between electricity and magnetism is obeyed and the magnetic monopoles had existed, there exists an interaction between two magnetic monopoles following a dual magnetic Coulomb's law,
\begin{equation}
F=\frac{1}{4\pi \mu _{0}}\frac{Q_{m1}Q_{m2}}{r^{2}}=k_{m}\frac{Q_{m1}Q_{m2}}{r^{2}}.
\end{equation}
Similarly the Planck magnetic monopole or Planck magnetic charge $Q_{mP}$ can be deduced,
\begin{equation}
k_{m}=1\cdot \frac{m_{P}l_{P}^{3}}{t_{P}^{2}Q_{mP}^{2}}=\frac{1}{4\pi \mu_{0}}=\frac{1}{4\pi \times 4\pi \times 10^{-7}N/A^{2}}.
\end{equation}
Previous Planck quantities substituted in, the Planck magnetic monopole definition and value follow in,
\begin{equation}
Q_{mP}=\sqrt{\frac{\hbar c}{k_{m}}}\approx 7.065751\times 10^{-16}Wb.
\end{equation}
Had the magnetic monopoles existed, the Planck magnetic monopole would be the signal value for them to play with gravity, and near or at least beyond which the validity of the perturbative quantum theory of magnetic monopoles would be spoiled.

The Planck electric current and Planck magnetic current are similarly defined limits for force laws in electromagnetism to avoid blundering into the Planck scale. One Ampere of current is such defined by combining Biot-Savat's Law and Lorentz force law that the magnetic force between a unit length section of two equal and parallel electric currents shall be some particular constant value.
\begin{equation}
F=\frac{\mu _{0}I^{2}L}{2\pi r}.
\end{equation}
In the Planck units, the Planck electric current follows,
\begin{equation}
[I]^{2}=\left[\frac{2\pi}{\mu_{0}}\right]\cdot [F] = \left[ \frac{c^2}{2k_{e}} \right] \cdot [F],
\end{equation}
\begin{eqnarray}
I=\sqrt{\frac{c^2}{2k_{e}}\frac{m_{P}l_{P}}{t_{P}^{2}}}=\sqrt{\frac{c^{6}}{2k_{e}G}}=\frac{Q_P}{\sqrt{2}t_P} \\
\approx 2.46002\times 10^{25}A.
\end{eqnarray}

The Maxwell equations and force laws involving magnetic monopoles can be derived using the duality between electricity and magnetism. An interaction between two separated parallel magnetic current sections shall follow a force law dual to that of the electricity in the above,
\begin{equation}
F=\frac{\epsilon _{0}I_{m}^{2}L}{2\pi r}.
\end{equation}
The Planck magnetic current is similarly defined,
\begin{eqnarray}
I_{m}=\sqrt{\frac{c^2}{2k_{m}}\frac{m_{P}l_{P}}{t_{P}^{2}}}=\sqrt{\frac{c^{6}}{2k_{m}G}}=\frac{Q_{mP}}{\sqrt{2}t_P}\\
\approx 9.26764\times 10^{27}Wb\cdot s^{-1}.
\end{eqnarray}

The above defined Planck parameters set the limit for the strength of various interactions accessible to perturbative quantum field theories. This becomes a constraint on the existence of magnetic monopoles.

\section{Constraints on Magnetic Monopoles by Running Dirac Dual Electric Charge}

As explained in Section II that the essential idea of the Planck mass/energy scale is the ultra strong and comparable strength of quantum effects and gravitational effects of physical processes occurring near the special energy scale. It is heuristic to see by directly plugging the definition of various Planck charges and Planck currents back into their respective force laws, that the strength of gravitation, the electric interaction and magnetic interaction under the situation bearing the name as Planck scale/Planck limit share common magnitude,
\begin{eqnarray}
F_P &&= G\frac{m_{P}^2}{l_{P}^{2}}=k_{e}\frac{Q_{P}^2}{l_{P}^{2}}=k_{m}\frac{Q_{mP}^2}{l_{P}^{2}} \nonumber\\
&&=\frac{\mu _{0}I_{P}^{2}l_P}{2\pi l_P}=\frac{\epsilon _{0}I_{mP}^{2}l_P}{2\pi l_P}= \frac{\hbar c}{\l_{P}^2} = \frac{c^4}{G} = m_{P}\frac{l_{P}}{t_{P}^{2}}.
\end{eqnarray}
Obviously we see the hint that all fundamental interactions under this situation require a unified treatment. All of the coupling constants, including the electric fine structure constant tend to unit value,
\begin{eqnarray}
\alpha_{G} = \frac{G}{\hbar c}\cdot m^{2}=\left( \frac{m}{m_{P}}\right) ^{2}\rightarrow &&1,\text{ }m\rightarrow m_{P}; \nonumber\\
\alpha = \frac{k_{e}}{\hbar c}\cdot e^{2}=\left( \frac{e}{Q_{P}}\right) ^{2}\rightarrow &&1,e\rightarrow Q_{P}; \nonumber\\
\alpha_{m} = \frac{k_{m}}{\hbar c}\cdot Q_{m}^{2}=\left( \frac{Q_{m}}{Q_{mP}} \right) ^{2}\rightarrow &&1,\text{ }Q_{m}\rightarrow Q_{mP}.
\end{eqnarray}

The running(growth or decrease) of the coupling constant of interactions with energy scale is a result of the renormalization group flow of our present perturbative quantum field theories. The gravitation, though very weak now, had been very strong interaction during the early universe when the average energy per particle was as high as the Planck mass/energy because then its coupling constant was near one, $\alpha_{G} \approx 1$. Depending on the behaviour of the theory, either at pretty low or pretty high energy scale, when the coupling constant approaches to one, the validity of the ordinary perturbative quantum field theory, which worked relying on the smallness of coupling constant, would be damaged.

The coupling constants being equal to one can be defined as the critical split between the strong coupling interactions and the weak coupling interactions. Of these the weak ones are the only ones which are manipulable by our familiar ordinary perturbative quantum field theories. Exploiting this idea as the constraints on the existence of magnetic monopoles, three consequences would immediately follow:

1) the corresponding energy scale when $\alpha_{m} \approx 1$ would become the fade-in/fade-out threshold of the perturbative magnetic monopoles;

2) the range of the value of the weak magnetic coupling constant is equivalent to the range of the magnitude of the magnetic charge of prehensible/catchable magnetic monopoles;

3) the constraints on magnetic monopole can be traded for that on the electric charge through an electric-magnetic duality carried out by the Dirac electric-magnetic charge quantization relation.

Note a relation between the face value of the Planck charges, the Planck mass, the Planck electric charge and the Planck magnetic monopole.
\begin{eqnarray}
Q_{P} &&= \sqrt{\frac{G}{k_{e}}}m_{P}, \nonumber\\
Q_{mP} &&= \sqrt{\frac{k_{e}}{k_{m}}}Q_{P} =\sqrt{\frac{k_{e}}{k_{m}}} \sqrt{\frac{G}{k_{e}}}m_{P} = \sqrt{\frac{G}{k_{m}}}m_{P}.
\end{eqnarray}

Now we are ready to exploit the implications of the Dirac electric-magnetic charge quantization relation,\cite{Dirac}
\begin{equation}
Q\cdot Q_{m}=2\pi \hbar \cdot n,\text{ }n\in Z.
\end{equation}
Firstly, it is interesting to see what value $n_{P}$ the quantum number $n$ should take if both the electric and magnetic charges take their Planck ceiling values,
\begin{equation}
Q_{P}\cdot Q_{mP}=2\pi \hbar \cdot n_{P},\text{ }n\in Z,
\end{equation}
\begin{equation}
n_{P}=\frac{Q_{P}\cdot Q_{mP}}{2\pi \hbar }=\frac{\sqrt{\frac{\hbar c}{k_{e}}} \cdot \sqrt{\frac{\hbar c}{k_{m}}}}{2\pi \hbar }=2
\end{equation}
Thus, $n_{P}=2$. It is such a surprisingly small value! Note also that,
\begin{equation}
Q_{mP} = 2 \cdot \frac {2 \pi \hbar}{Q_{P}} = \frac {4\pi \hbar}{Q_{P}} \label{Both=Placnk}
\end{equation}

The smallness of $n_P=2$ is conceptually very impressive. Planck charges are huge values which the electric charge and magnetic monopole values should both not reach, in order to stay within the region of the current perturbative quantum field theory. It then squeezes the survival possibility for magnetic monopoles to only two practical choices:

1) $n=0$, $Q\cdot Q_{m}=2\pi \hbar \cdot n = 0 \Rightarrow Q_{m}=0$,

No magnetic monopoles can exist at all, since we do live with electric charges every day, $Q_{mP} \neq 0$;

2) $n=1$, $Q\cdot Q_{m}=2\pi \hbar$,

This is the only survival possibility allowable by the Dirac electric-magnetic charge quantization relation combined with the requirement for the Planck limits not be violated. Here we see such a simplistic but powerful argument for the scarcity of the magnetic monopoles in the universe.

Knowing that $n=1$, we have,
\begin{equation}
Q_{m}=\frac{2\pi \hbar }{Q}\cdot n\Big|(n=1) \Rightarrow Q_{m}=\frac{2\pi \hbar }{Q}. \label{n=1}
\end{equation}
The value of the magnetic monopole is inversely proportional to the value of the electric charge. The tiny electric charge $Q\rightarrow e$ at the low end of energy scale maps to a magnetic monopole value greater than the Planck magnetic monopole, which is forbidden, or unobservable.
\begin{equation}
Q_m=\frac{2\pi \hbar }{e}\approx \frac{2\pi \hbar }{(Q_{P}/11.706)} = 5.853 Q_{mP} \textgreater Q_{mP}
\end{equation}
The tendency of the behaviour of the renormalization group equations of the electric and magnetic coupling constants should be opposite. The electric charge value would grow with the increasing energy scale, but the magnetic monopole decreases with it. Dividing eq.(\ref{Both=Placnk}) by eq.(\ref{n=1}) then squaring, and note that $\alpha =\left( \frac{Q}{Q_{P}} \right) ^{2}$ and $\alpha_{m} =\left( \frac{Q_{m}}{Q_{mP}} \right) ^{2}$, we the a relation between the coupling electric and magnetic constants,
\begin{equation}
\alpha \cdot \alpha_m=\frac{1}{4}.
\end{equation}
Also note, $Q < Q_{P} \Leftrightarrow \alpha < 1$. It takes $\alpha \geq \frac{1}{4}$ for $\alpha_m \leq 1,\, Q_m \leq Q_{mP}$. This happens at some pretty high energy scale. We need to go to pretty high energy scales when $\alpha > \frac{1}{4}$ so that $\alpha_m < 1,\, Q_m < Q_{mP}$ and then magnetic monopoles can become observable objects handleable by perturbative quantum field theories.

The magnetic monopole approaches to a minimum value limit when $Q \rightarrow Q_{P}$,
\begin{eqnarray}
Min\left[Q_{m}\right] \rightarrow \frac{2\pi \hbar }{Q_{P}}=\frac{1}{2}\left(\frac{4\pi \hbar }{Q_{P}}\right) = \frac{1}{2}Q_{mP} \\
\approx 3.532876 \times 10^{-16}Wb.
\end{eqnarray}
Had magnetic monopoles really existed, this should be the minimum effect they will bring to physical experiments. Even though this value does not protrude across the maximum Planck limit, it is not too much smaller than the limit. We are convinced that magnetic monopoles would never be easily handleable object.

\section{Model Dependency of Renormalization Group Flow}

We see from the last section that, the Dirac duality between the electric and magnetic charges allows for the treatment of magnetic charges to be traded with electric charge. We know that, for $\alpha_m \leq 1,\, Q_m \leq Q_{mP}$, it takes $\alpha \geq \frac{1}{4}$. Of course $\alpha < 1$, we thus require,
\begin{equation}
 \frac{1}{4} \leq \alpha < 1.
\end{equation}
We call this value range in terms of the electric fine structure constant the hermitic Shangri-La for magnetic monopoles, a remote but dreamy paradise physicists have been seeking for centuries where the magnetic monopoles live their fairytale life. There and only there is where physicists should go and search.

In terms of the energy scales, it depends on the different behaviour of specific renormalization group equation, or the so-called Callan-Symanzik equation of different models. For the quantum electrodynamics(QED), the electric fine structure constant is the only coupling constant. The running rate of charge with energy scale $\mu$, denoted by the so-called $\beta$ function, at the one-loop level, gives, in natural units,\cite{Peskin}
\begin{equation}
\beta(e) = \frac{de}{d\mu} = \frac{e^3}{12 \pi^2}.
\end{equation}
Solving it gives the running of the electric coupling constant with energy scale, or equivalently the momentum transfer $q$ of physical processes,
\begin{equation}
\bar{\alpha}(q)=\frac{\alpha}{1-(\frac{\alpha}{3\pi}){\rm log}(\frac{q^2}{Am^2})},
\end{equation}
where $A=exp(\frac{5}{3})$, $m$ electron rest mass. It can be seen that the coupling constant increases with energy scale, or the increasing momentum transfer $q^2$. By a rough numeric estimate, we see that if $\bar{\alpha}=0.25$, $\Rightarrow$ super large energy scale, much greater than Planck mass/energy. It is a negative hint for the existence of magnetic monopoles.

In the Weinberg-Salam $SU(2)\times U(1)$ model of electroweak unification of the particle standard model, the electric interaction is not independent. Then the electric charge $e$ is the combination of two gauge couplings,
\begin{equation}
e=\frac{gg'}{\sqrt{g^2+g'^2}}=\frac{g_1g_2}{\sqrt{g_1^2+g_2^2}}.
\end{equation}
Defining $\alpha_i=\frac{g_i^2}{4\pi}$, we have,
\begin{equation}
\alpha=\frac{\alpha_1\cdot\alpha_2}{\alpha_1+\alpha_2}.
\end{equation}
The running of these coupling constants are the following,\cite{Mohapatra}
\begin{eqnarray}
&& \frac{d\alpha_1}{dt}=-\frac{\alpha_1^2}{2\pi}\cdot\frac{2}{3}N_f,\nonumber\\
&& \frac{d\alpha_2}{dt}=-\frac{\alpha_2^2}{2\pi}\cdot\left(\frac{22}{3} - \frac{2}{3}N_f\right),\nonumber\\
&& t\equiv log \mu.
\end{eqnarray}
Here $N_f$ is the number of the flavor of quarks, which is usually taken to be six, though it is possible to be augmented in models with more than three generations of fermions. The initial values of those first order differential equations, the value of the corresponding coupling constants at low energy scales have already been measured by particle experiments. Further analysis of the running curves can be executed by numeric calculations.

Same tendency can be obtained with analysis based on electroweak theory or different GUT models. Different particle theory models can give somewhat different running curves, differing slightly in the prediction of specific energy ranges. Whether or not the electric coupling constant can go beyond the value $\alpha=\frac{1}{4}$ before the Planck mass/energy, and at what specific energy scale does this, it is upto different models of particle physics. The above argument provides intriguing qualitative direction to the efforts in search of magnetic monopoles.

\section{Summary and Conclusion}
%\begin{center}
%\textbf{V. Summary and Conclusion}
%\end{center}

Near a high energy scale characterized by the Planck mass, the strength of gravity ceases being weak and demands a unified treatment with other fundamental interactions. The running tendency of the coupling constants caused by the renormalization group flow in perturbative quantum field theories demands that limits be specified for various interactions to stop them from running into the region where the perturbative quantum field theories themselves collapse. These limits become the conceptual foundation for the definition of Planck charges. They together form a set of boundary markers signaling the buffer zone preventing the ordinary physics below Planck scale from running into the sphere of high energy quantum gravity physics.

When applied to magnetic monopole case obeying Dirac's electric-magnetic charge quantization relation, the maximum allowed Planck electric charge would mean a lower bound to the value of magnetic monopole as half of the Planck magnetic monopole. As the electric charge and magnetic charge evolve inversely with energy scale, the electric charge can not go lower than half of the Planck electric charge for the magnetic charge not to touch the Planck magnetic monopole ceiling. Thus in terms of the value of the electric fine structure constant, the range $\frac{1}{4} \textless \alpha \textless 1$, is the survival range for magnetic monopoles. Oue result confirms a similar conclusion in literature in which the authors refered only to concept of the limit of perturbative theories, but did not refer to the concept of Planck charge limit, those two ideas we have shown to be equivalent.\cite{LaperashviliNielsen} We claim to have pointed out the location of the hermitic Shangri-La where the magnetic monopole are inhabiting, and in general the way to approach it. The specific range of the energy scales we recommend to go for search of the magnetic monopoles have been shown to be model dependent. It can provide discriminative directions to the century-long project in search of magnetic monopoles.

At last we note that the curve of the running coupling constant for magnetic monopole resembles that of quantum chromodynamics with asymptotic freedom. It implies the possibility of borrowing the concepts of quark confinement as one way of explanation for the difficulty of finding magnetic monopoles. It might be conjectured that the strong coupling confined magnetic monopole balls might be a possible component for cosmological dark matter in which direction some research attempts might be considered relevant.\cite{XingangChen}

%\begin{acknowledgments}
%\begin{center}
\textbf{Acknowledgments}
%\end{center}

Authors are grateful to Prof. Anzhong Wang and Prof. Gerald Cleaver for useful discussions. This work was supported in part by the National Natural Science Foundation of China (NNSFC) (Grants No. 11275017 and No. 11173028).
%\end{acknowledgments}

Author email addresses:\\
$^{1}$yanbin\_deng@baylor.edu\\
$^{2}$cyhuang@purdue.edu\\
$^{3}$ychuang@bjut.edu.cn

\end{document}